\documentclass[12pt]{article}
\usepackage{amsmath,amssymb}
\usepackage{layout}
\usepackage{latexsym,graphicx}
\usepackage[cp1251]{inputenc}
\usepackage{curves,eepic}

\usepackage{floatflt}

\usepackage{wrapfig}
\usepackage{amsmath,amssymb}

\newcommand{\be}{\begin{equation}}
\newcommand{\ee}{\end{equation}}
\newcommand{\bes}{\begin{subequations}}
\newcommand{\ees}{\end{subequations}}
\newcommand{\bea}{\begin{eqnarray}}
\newcommand{\eea}{\end{eqnarray}}
\newcommand{\bear}{\begin{equation}\begin{array}}
\newcommand{\eear}[1]{\end{array}\label{#1}\end{equation}}
\def\ba{$$\begin{array}}
\def\ea{\end{array}$$}
\def\bra{$\begin{array}}
 \def\era{\end{array}$}

\def\cl{\centerline}

\newsavebox{\fmbox}

{\end{list}}
\newcounter{enumct}

\newcommand{\bu}{$\bullet$\ }

\begin{document}
\date{}

\title{Subcritical Fission Reactor  Based on
Linear Collider}

\author{ I.~F.~Ginzburg\\
{\small\it Sobolev Institute of Mathematics}\\ {\small\it
and Novosibirsk State University,}\\ {\small\it
Novosibirsk, 630090, Russia}}

\maketitle

\begin{abstract}
The beams of Linear Collider after main collision can be
utilized to build an accelerator--driven sub--critical
reactor.
\end{abstract}

$\blacksquare$ The project of Linear Collider (LC) contains
one essential element that is not present in other
colliders. Here each electron (or positron or photon) bunch
will be used only once, and physical collision leave two
very dense and strongly collimated  beams of high energy
electrons or/and photons with precisely known time
structure. We consider, for definiteness, electron beam
parameters of the TESLA project \cite{TESLA}
  \bear{c}
particle\;\; energy\;\; E_e=250\;GeV, \\ number\;\;
of\;\;electrons\;\; per\;\; second\;\; N_e=2.7\cdot
10^{14}/s,\\ mean\;\;beam\;\; power\;\;P_b\approx 11\;MWt,\\
transverse\;\; size\;\; and\;\; angular\;\;
spread\;\;negligible.
 \eear{beampar}
In the Photon Collider mode the used beams contain photons,
electrons and positrons. They are not monochromatic but
have the same characteristic particle energy (with large
low energy tail) and the same mean power.

The problem, how to deal with this powerful beam dump, is
under intensive discussion.

Main discussed variant is to destruct these used beams with
minimal radioactive pollution (see e.~g.~\cite{TESLA}). It
looks natural also to use these once--used beams for fixed
target experiments with unprecedented precision.

Recently we suggested to utilize these used beams  to
initiate work of subcritical fission reactor and to
construct neutrino factory \cite{LCWS05}. Here we present
estimates for one of these options. Real choice and
optimization of parameters should be the subject of detail
subsequent studies.

$\blacksquare$ The idea to work with sub--critical nuclear
reactor, initiated by proton or electron beam, for
foolproof production of energy and/or cleaning of nuclear
pollution is well known (see e.~g.~\cite{Rubbia}). Here
proton or electron beam with particle energy of about 1 GeV
is supposed to produce neutrons in the cascades within body
of reactor. The problem here is in obtaining necessary beam
power $P_b\ge 5$~MW.

According to \eqref{beampar}, each used beam of LC is two
times more powerful than necessary for this reactor but
electron energies are two orders larger. In the suitable
target this particle energy can be transmitted to low
energy particles to initiate fission process in reactor.

\cl{\bf Qualitative description}

\bu The first redistribution of beam energy to a large
number of "working" electrons and photons can be realized
in special {\it degrader} -- e.g., 0.5~m water  pipe with
the radius of a few cm. (Water should rotate to prevent
vapour explosion.) After passing the degrader, particles
with mean energy in hundreds MeV penetrate into the main
body of reactor filled with uranium or thorium. After the
photons reach the energy of about 10 MeV  in the
electromagnetic cascades, they get absorbed by nuclei (in
giant resonance), producing neutrons.

The scheme of proper reactor is a subject of separate study
of reactor specialists.

\bu To realize this reactor, the crab crossing scheme for
main collision may be preferable to place the reactor away
from accelerator beam. We assume for definiteness the crab
crossing angle of 15 mrad.

One or two sub--critical reactors can be situated at about
500 m from collision point, at about 7 m from accelerating
channel providing good  protection of collider beam pipe.
(Considered used beam should move towards the reactor
through low pressure gas after it passes some window
protecting high vacuum of collider.)

\bu The obtained accelerator--driven foolproof
sub--critical reactor can be used for energy generation and
extra nuclear pollution cleaning. The economical problems
are beyond this proposal.\\

I am thankful to   V.~Saveliev, A.~Sessler, A.~Skrinsky,
V.~Telnov,  M.~Zolotarev for comments and new information.
This work is supported by grants RFBR 05-02-16211,
NSh-2339.2003.2.

\end{document}